\documentclass[aps,epsf,rotate,citesort,preprint]{revtex4}

\usepackage{graphicx}
\usepackage{times}
\usepackage{natbib}

\begin{document}


\centerline{\large \bf Comment on \textit{The origin of bursts and heavy tails in human dynamics}}
\vspace{2.0cm}

In a recent letter, Barab\'asi claims that the dynamics of a number of
human activities are scale-free~\cite{barabasi05}.  He specifically
reports that the probability distribution of time intervals $\tau$
between consecutive e-mails sent by a single user and time delays for
e-mail replies follow a power-law, $P(\tau)\approx\tau^{-\alpha}$ with
$\alpha \simeq 1$, and proposes a priority-queuing process as an
explanation of the ``bursty'' nature of human activity.  Here, we
quantitatively demonstrate that the reported power-law distributions
are solely an artifact of the analysis of the empirical data and that
the proposed model is not representative of e-mail communication
patterns.

Barab\'asi analyzed the email communication patterns of a subset of
users \cite{amaral05a} in a database containing the email usage
records of 3188 individuals using a university e-mail server over an
83-day period~\cite{eckmann04}.  Upon examining the same data, we find
a number of significant deficiencies in his analysis.  These
deficiencies were communicated to Barab\'asi well in advance of
publication~\cite{amaral05}.  For example, even though the data have a
resolution of one second, the statistical analysis reported in Fig.~2
of Ref.~\cite{barabasi05} indicates that the most frequent time
interval between consecutive e-mails sent by the same user occurs for
time intervals {\it smaller\/} than one second.  Even more
surprisingly, the user considered in Fig.~2 of Ref.~\cite{barabasi05}
appears to respond to e-mails most frequently for times smaller than
five seconds. We verified that such time intervals
are too short to permit a person to write and send consecutive
e-mails, much less read, write, and reply to an e-mail.

Unfortunately, these are not the only problems with the claims of
Ref.~\cite{barabasi05}.  Barab\'asi claims that the time series of the
typical user is well-described by a power-law distribution with an
exponent $\alpha \simeq 1$.  This claim is revised in more recent
work, which suggests that the power-law is modified by an exponential
truncation~\cite{vazquez05,vazquez05a}.  Our own analysis of the same
empirical data used in Ref.~\cite{barabasi05} suggests that a
log-normal distribution provides a significantly better description of
the data.

\begin{figure}
\centerline{\includegraphics*[width=0.8\columnwidth]{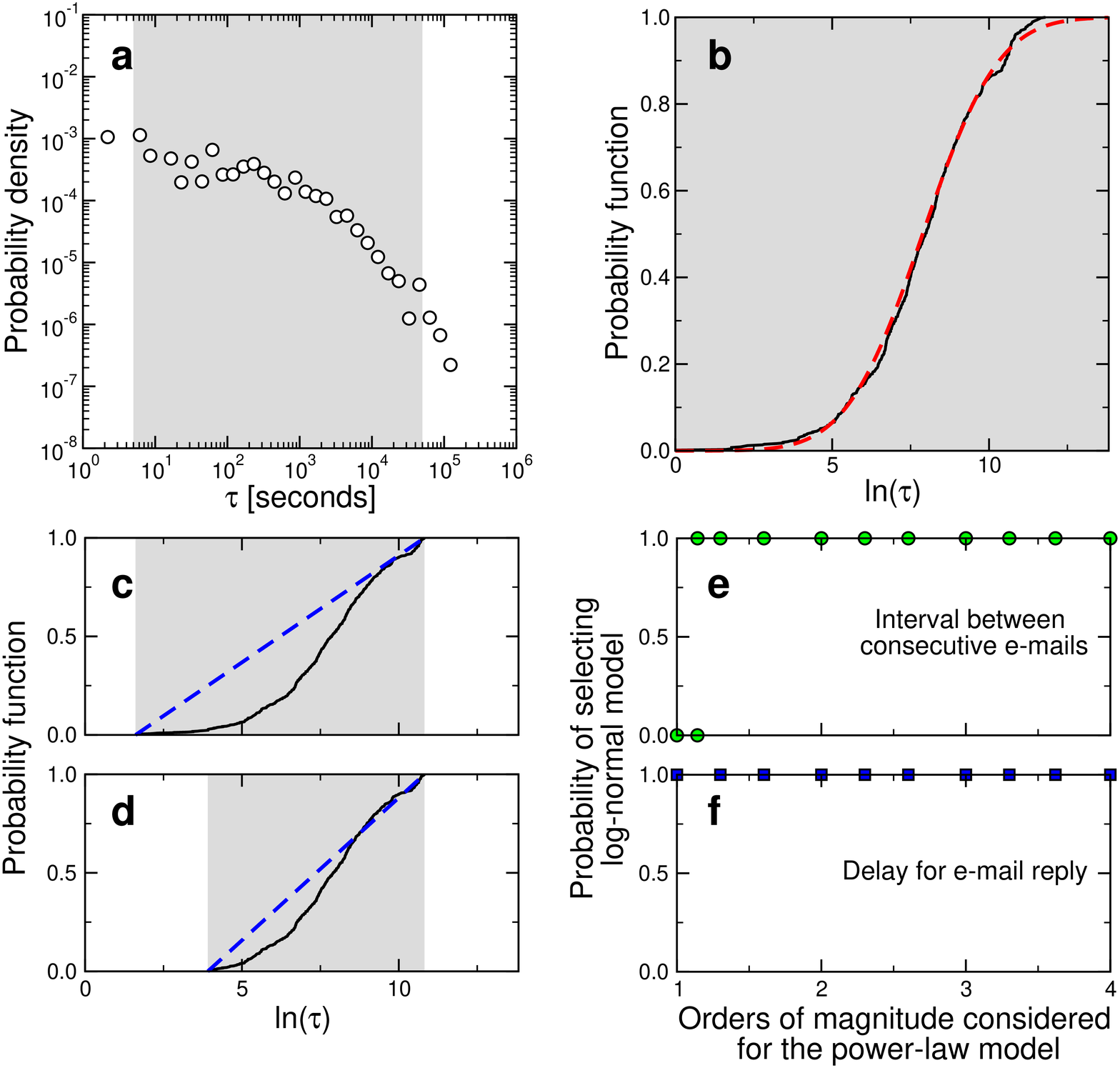}}
\vspace*{-0.3cm}
\renewcommand{\baselinestretch}{1.0}
\caption{Bayesian model selection protocol for comparing the power-law
  with exponent $\alpha=1$ and the log-normal models. {\bf a},
  Probability density of the time intervals $\tau$ between consecutive
  e-mails sent for a user who sent approximately 8 e-mails per day
  during the study period~\cite{eckmann04}.  The region shaded in grey
  $\tau \in [5,50000]$ corresponds to the range reported
  in~\cite{barabasi05,vazquez05,vazquez05a} to be well-approximated by
  a power-law with exponent $\alpha=1$.  {\bf b}, To investigate the
  validity of the log-normal hypothesis, we take advantage of the fact
  that the probability function $P(s)$ of the logarithm of the time
  interval $s = \ln(\tau)$ should be Gaussian.  It is visually
  apparent that the probability function for the user considered in
  {\bf a} (black curve) is well-described by a Gaussian (dashed red
  curve).  {\bf c}, To investigate the validity of the power-law with
  an exponent $\alpha=1$, we can perform a similar transformation and
  analysis.  If the data found in the grey shaded region of {\bf a}
  were well-described by a power-law with exponent $\alpha=1$, the
  $P(s)$ would be linear with slope $(s_{max} - s_{min})^{-1}$.  It is
  visually apparent that the probability function for the data in the
  range $\ln(5) \le s \le \ln(50000)$ is \emph{not} linear.  {\bf d},
  Even when the considered data range is reduced one order of
  magnitude to $\ln(50) \le s \le \ln(50000)$ the data are
  \emph{still} not linear.  {\bf e}, We conduct Bayesian model
  selection analysis between the two candidate models for all 1202
  users who sent more than 10 messages during the study
  period~\cite{eckmann04}.  When considering the time interval between
  consecutive emails, we find that one would effectively always select
  the log-normal model over the power-law model, except when the
  power-law model is used for \emph{one order of magnitude or less} in
  the time domain. {\bf f}, We perform a similar analysis for all 760
  users who reply to at least 10 messages during the study
  period~\cite{eckmann04}.  When considering the time delay for e-mail
  replies, we find that one would always select the log-normal model
  over the power-law model, even when considering \emph{just one order
  of magnitude} of the data.  It is critical to note that in the
  analysis of both {\bf e} and {\bf f} we consider \emph{all} user
  data for the log-normal model in contrast to the reduced range
  considered for the power-law model.}
\label{distributions}
\end{figure}

Our hypothesis of a log-normal distribution may also be more
appropriate for describing the activity of users that rely on e-mail
for daily communication for the following reasons.  To our knowledge,
there are no studies reporting that a real-world process is
well-described by a power-law with an exponent $\alpha \simeq 1$.  An
\emph{apparent} scaling exponent $\alpha \simeq 1$ and concave
curvature depicted in Fig. 2A of Ref.~\cite{barabasi05} and numerous
figures of Ref.~\cite{vazquez05a} are, however, characteristics of a
log-normal distribution, which are representative of many real-world
processes~\cite{mitzenmacher04}.  Log-normal distributions are easily
identifiable by examining the probability density of $s =
\ln(\tau)$.  Under this transformation, a power-law with exponent
$\alpha \simeq 1$ would be a uniform distribution of $s$.  It is
visually apparent that $P(s)$ is not uniform, but rather Gaussian in
form (see \textit{Supplementary Information}).  The Gaussian form of
this distribution suggests that users send consecutive e-mails with a
characteristic time ($\tau \approx 45$ minutes for the user in
Fig.~\ref{distributions}) as opposed to Barab\'asi's contention that
users send e-mails without a characteristic scale.

We conduct a Bayesian model selection analysis to decide between the
two proposed descriptions of the data~\cite{mood74,bernardo00}.  As in
Ref.~\cite{barabasi05}, we analyze both the time intervals between
consecutive e-mails sent and the time required to reply to an e-mail.
To be as considerate as possible with the analysis of
Ref.~\cite{barabasi05}, we assume prior probabilities of 90\% for the
truncated power-law model and 10\% for the log-normal model.  A more
stringent comparison would give each model equal likelihood of
describing the data in the absence of additional information.
Furthermore, we restrict the time domain for our analysis of the
power-law whereas we consider the \emph{entire} time domain for our
analysis of the log-normal distribution.  We find that the posterior
probability of the log-normal description being correct is {\it
indistinguishable from one} within the computer's numerical precision.

Additionally, we calculate posterior probabilities as a
function of the magnitude of the power-law domain.  As we show in
Fig.~\ref{distributions}, the log-normal distribution provides a
better description of the data than a power-law except when less than
one order of magnitude is considered for the analysis of the
power-law (see Fig.~\ref{distributions} and the
\textit{Supplementary Information} for full details of this analysis).

We next discuss the priority-queuing model which reportedly explains
the mechanism behind the reply times in e-mail communication.  Before
addressing the details of the model, however, we would like to
emphasize that the model predicts a power-law for the distribution of
response time delays, \emph{not} the empirically observed log-normal
distribution.  These predictions of the model are supported by
recent analytical work by V\'azquez et
al.~\cite{vazquez05,vazquez05a}.  As we demonstrate above, that
prediction is \emph{not} supported by the data. 


The priority-queuing model is not only unrealistic in its prediction
of the functional form of the distribution of time delays for e-mail
responses.  After an initial transient period, new tasks in the model
are typically executed immediately after arrival, resulting in a
pronounced peak at $\tau=1$; these tasks are said to represent e-mails
which are either immediately replied to or deleted~\cite{barabasi05}.
In the case that reportedly best captures human
dynamics~\cite{barabasi05}, that is, when $\epsilon \rightarrow 0$,
where $\epsilon$ is the probability of executing a randomly-selected
task instead of the highest-priority task, this peak contains $99.9$\%
of the tasks handled by the user---an unrealistic scenario.  Moreover,
upon reaching steady state, the distribution of task priorities on the
queue converges to a uniform distribution in the interval
$[0,\epsilon]$.  The model thus predicts that the typical e-mail user
has a queue filled with \emph{extremely} low priority tasks and
consequently performs all new incoming tasks immediately upon arrival.
This situation is also not representative of \emph{typical} human
behavior.

E-mail communication patterns are a valuable proxy for the study of human 
behavior and decision-making.  The idea of humans relying solely on a
priority-queuing procedure~\cite{gross98} to manage their complex
activity is interesting.  Unfortunately, we find that even though 
Ref.~\cite{barabasi05} is quite stimulating, \emph{none} of the results it 
reports hold upon further inspection.

\vspace*{1cm}
\noindent
{\bf Daniel B. Stouffer, R. Dean Malmgren, Lu\'{\i}s A. Nunes Amaral}
\\
\noindent
Department of Chemical and Biological Engineering, Northwestern
University,\\ Evanston, IL 60201, USA \\


\newpage


\end{document}